\RequirePackage{lineno}
\documentclass[aps,prl,twocolumn,showpacs,groupedaddress,superscriptaddress]{revtex4}
\usepackage{graphicx}  
\usepackage{dcolumn}   
\usepackage{bm}        
\usepackage{amssymb}   
\usepackage{amsmath}

\def\Etmiss{\hbox{$\rlap{\kern0.25em/}E_T$}}
 
\begin{document}


\title{Measurement of the $\boldsymbol{ZZ}$ production cross section in $\boldsymbol{p\bar{p}}$ collisions at $\boldsymbol{\sqrt{s}=}$1.96~TeV}

%

\affiliation{Universidad de Buenos Aires, Buenos Aires, Argentina}
\affiliation{LAFEX, Centro Brasileiro de Pesquisas F{\'\i}sicas, Rio de Janeiro, Brazil}
\affiliation{Universidade do Estado do Rio de Janeiro, Rio de Janeiro, Brazil}
\affiliation{Universidade Federal do ABC, Santo Andr\'e, Brazil}
\affiliation{Instituto de F\'{\i}sica Te\'orica, Universidade Estadual Paulista, S\~ao Paulo, Brazil}
\affiliation{Simon Fraser University, Vancouver, British Columbia, and York University, Toronto, Ontario, Canada}
\affiliation{University of Science and Technology of China, Hefei, People's Republic of China}
\affiliation{Universidad de los Andes, Bogot\'{a}, Colombia}
\affiliation{Charles University, Faculty of Mathematics and Physics, Center for Particle Physics, Prague, Czech Republic}
\affiliation{Czech Technical University in Prague, Prague, Czech Republic}
\affiliation{Center for Particle Physics, Institute of Physics, Academy of Sciences of the Czech Republic, Prague, Czech Republic}
\affiliation{Universidad San Francisco de Quito, Quito, Ecuador}
\affiliation{LPC, Universit\'e Blaise Pascal, CNRS/IN2P3, Clermont, France}
\affiliation{LPSC, Universit\'e Joseph Fourier Grenoble 1, CNRS/IN2P3, Institut National Polytechnique de Grenoble, Grenoble, France}
\affiliation{CPPM, Aix-Marseille Universit\'e, CNRS/IN2P3, Marseille, France}
\affiliation{LAL, Universit\'e Paris-Sud, CNRS/IN2P3, Orsay, France}
\affiliation{LPNHE, Universit\'es Paris VI and VII, CNRS/IN2P3, Paris, France}
\affiliation{CEA, Irfu, SPP, Saclay, France}
\affiliation{IPHC, Universit\'e de Strasbourg, CNRS/IN2P3, Strasbourg, France}
\affiliation{IPNL, Universit\'e Lyon 1, CNRS/IN2P3, Villeurbanne, France and Universit\'e de Lyon, Lyon, France}
\affiliation{III. Physikalisches Institut A, RWTH Aachen University, Aachen, Germany}
\affiliation{Physikalisches Institut, Universit{\"a}t Freiburg, Freiburg, Germany}
\affiliation{II. Physikalisches Institut, Georg-August-Universit{\"a}t G\"ottingen, G\"ottingen, Germany}
\affiliation{Institut f{\"u}r Physik, Universit{\"a}t Mainz, Mainz, Germany}
\affiliation{Ludwig-Maximilians-Universit{\"a}t M{\"u}nchen, M{\"u}nchen, Germany}
\affiliation{Fachbereich Physik, Bergische Universit{\"a}t Wuppertal, Wuppertal, Germany}
\affiliation{Panjab University, Chandigarh, India}
\affiliation{Delhi University, Delhi, India}
\affiliation{Tata Institute of Fundamental Research, Mumbai, India}
\affiliation{University College Dublin, Dublin, Ireland}
\affiliation{Korea Detector Laboratory, Korea University, Seoul, Korea}
\affiliation{CINVESTAV, Mexico City, Mexico}
\affiliation{FOM-Institute NIKHEF and University of Amsterdam/NIKHEF, Amsterdam, The Netherlands}
\affiliation{Radboud University Nijmegen/NIKHEF, Nijmegen, The Netherlands}
\affiliation{Joint Institute for Nuclear Research, Dubna, Russia}
\affiliation{Institute for Theoretical and Experimental Physics, Moscow, Russia}
\affiliation{Moscow State University, Moscow, Russia}
\affiliation{Institute for High Energy Physics, Protvino, Russia}
\affiliation{Petersburg Nuclear Physics Institute, St. Petersburg, Russia}
\affiliation{Instituci\'{o} Catalana de Recerca i Estudis Avan\c{c}ats (ICREA) and Institut de F\'{i}sica d'Altes Energies (IFAE), Barcelona, Spain}
\affiliation{Stockholm University, Stockholm and Uppsala University, Uppsala, Sweden}
\affiliation{Lancaster University, Lancaster LA1 4YB, United Kingdom}
\affiliation{Imperial College London, London SW7 2AZ, United Kingdom}
\affiliation{The University of Manchester, Manchester M13 9PL, United Kingdom}
\affiliation{University of Arizona, Tucson, Arizona 85721, USA}
\affiliation{University of California Riverside, Riverside, California 92521, USA}
\affiliation{Florida State University, Tallahassee, Florida 32306, USA}
\affiliation{Fermi National Accelerator Laboratory, Batavia, Illinois 60510, USA}
\affiliation{University of Illinois at Chicago, Chicago, Illinois 60607, USA}
\affiliation{Northern Illinois University, DeKalb, Illinois 60115, USA}
\affiliation{Northwestern University, Evanston, Illinois 60208, USA}
\affiliation{Indiana University, Bloomington, Indiana 47405, USA}
\affiliation{Purdue University Calumet, Hammond, Indiana 46323, USA}
\affiliation{University of Notre Dame, Notre Dame, Indiana 46556, USA}
\affiliation{Iowa State University, Ames, Iowa 50011, USA}
\affiliation{University of Kansas, Lawrence, Kansas 66045, USA}
\affiliation{Kansas State University, Manhattan, Kansas 66506, USA}
\affiliation{Louisiana Tech University, Ruston, Louisiana 71272, USA}
\affiliation{Boston University, Boston, Massachusetts 02215, USA}
\affiliation{Northeastern University, Boston, Massachusetts 02115, USA}
\affiliation{University of Michigan, Ann Arbor, Michigan 48109, USA}
\affiliation{Michigan State University, East Lansing, Michigan 48824, USA}
\affiliation{University of Mississippi, University, Mississippi 38677, USA}
\affiliation{University of Nebraska, Lincoln, Nebraska 68588, USA}
\affiliation{Rutgers University, Piscataway, New Jersey 08855, USA}
\affiliation{Princeton University, Princeton, New Jersey 08544, USA}
\affiliation{State University of New York, Buffalo, New York 14260, USA}
\affiliation{Columbia University, New York, New York 10027, USA}
\affiliation{University of Rochester, Rochester, New York 14627, USA}
\affiliation{State University of New York, Stony Brook, New York 11794, USA}
\affiliation{Brookhaven National Laboratory, Upton, New York 11973, USA}
\affiliation{Langston University, Langston, Oklahoma 73050, USA}
\affiliation{University of Oklahoma, Norman, Oklahoma 73019, USA}
\affiliation{Oklahoma State University, Stillwater, Oklahoma 74078, USA}
\affiliation{Brown University, Providence, Rhode Island 02912, USA}
\affiliation{University of Texas, Arlington, Texas 76019, USA}
\affiliation{Southern Methodist University, Dallas, Texas 75275, USA}
\affiliation{Rice University, Houston, Texas 77005, USA}
\affiliation{University of Virginia, Charlottesville, Virginia 22901, USA}
\affiliation{University of Washington, Seattle, Washington 98195, USA}
\author{V.M.~Abazov} \affiliation{Joint Institute for Nuclear Research, Dubna, Russia}
\author{B.~Abbott} \affiliation{University of Oklahoma, Norman, Oklahoma 73019, USA}
\author{B.S.~Acharya} \affiliation{Tata Institute of Fundamental Research, Mumbai, India}
\author{M.~Adams} \affiliation{University of Illinois at Chicago, Chicago, Illinois 60607, USA}
\author{T.~Adams} \affiliation{Florida State University, Tallahassee, Florida 32306, USA}
\author{G.D.~Alexeev} \affiliation{Joint Institute for Nuclear Research, Dubna, Russia}
\author{G.~Alkhazov} \affiliation{Petersburg Nuclear Physics Institute, St. Petersburg, Russia}
\author{A.~Alton$^{a}$} \affiliation{University of Michigan, Ann Arbor, Michigan 48109, USA}
\author{G.~Alverson} \affiliation{Northeastern University, Boston, Massachusetts 02115, USA}
\author{G.A.~Alves} \affiliation{LAFEX, Centro Brasileiro de Pesquisas F{\'\i}sicas, Rio de Janeiro, Brazil}
% ...
%
%
\collaboration{The D0 Collaboration\footnote{with visitors from
$^{a}$Augustana College, Sioux Falls, SD, USA,
$^{b}$The University of Liverpool, Liverpool, UK,
$^{c}$SLAC, Menlo Park, CA, USA,
$^{d}$University College London, London, UK,
$^{e}$Centro de Investigacion en Computacion - IPN, Mexico City, Mexico,
$^{f}$ECFM, Universidad Autonoma de Sinaloa, Culiac\'an, Mexico,
and 
$^{g}$Universit{\"a}t Bern, Bern, Switzerland.
$^{h}$Visitor from Bradley University, Peoria, IL, USA.
}} \noaffiliation
\vskip 0.25cm
\date{April 14, 2011}


\begin{abstract}

We present a new measurement of the production cross section $\sigma(p\bar{p} \rightarrow ZZ)$ at a center-of-mass energy $\sqrt{s}=1.96$~TeV, obtained from the analysis of the four charged lepton final state $\ell^{+}\ell^{-}\ell^{'+} \ell^{'-}$ ($\ell$, $\ell^{'}$ = $e$ or $\mu$).  We observe ten candidate events with an expected background of $0.37 \pm 0.13$ events.  The measured cross section $\sigma(p\bar{p} \to ZZ)= 1.26^{+0.47}_{-0.37}~\mathrm{(stat)} \pm 0.14~\mathrm{(syst)}$~pb is in agreement with NLO QCD predictions.  This result is combined with a previous result from the $ZZ\to\ell^{+}\ell^{-}\nu\bar{\nu}$ channel resulting in a combined cross section of $\sigma(p\bar{p} \to ZZ)= 1.40^{+0.43}_{-0.37}~\mathrm{(stat)} \pm 0.14~\mathrm{(syst)}$~pb.     


\end{abstract}

 
\maketitle

Studies of the pair production of electroweak gauge bosons provide
an important test of electroweak theory predictions.  The production of pairs of $Z/\gamma^*$ bosons has the smallest cross sections for any standard model (SM) diboson process not involving the Higgs boson.  The next-to-leading order (NLO) SM prediction for the $Z/\gamma^*Z/\gamma^*$ production cross section in $p\bar{p}$ collisions at the Fermilab Tevatron Collider at $\sqrt{s}=1.96$~TeV is $\sigma(p\bar{p} \to Z/\gamma^*Z/\gamma^*)=1.4 \pm 0.1$~pb~\cite{Campbell:1999ah}.  This cross section is evaluated in a high mass region where the masses of $(Z/\gamma^*)_{1}$ and $(Z/\gamma^*)_{2}$ are greater than 70 GeV and 50 GeV, respectively.  A correction factor of 0.93, derived using \textsc{pythia}~\cite{pythia}, is used to convert the measured cross section for $Z/\gamma^*Z/\gamma^*$ into that for $ZZ$ production.  Studies of this process are important not only to further test the SM, but also for Higgs boson searches.  Specifically, if the Higgs boson has a mass greater than 180 GeV, it will have a significant branching fraction into $Z$ boson pairs.  Thus, in that context, SM $Z/\gamma^*Z/\gamma^*$ production is an important background to Higgs boson searches.  Beyond the Higgs sector,  the observation of an unexpectedly high cross section could indicate the presence of anomalous $ZZZ$ or $ZZ\gamma$ couplings~\cite{zz-theory} or the existence of extra dimensions~\cite{ed-zz} or exotic particles.


Previous investigations of $Z/\gamma^*Z/\gamma^*$ production have been performed both at the
Fermilab Tevatron $p\bar{p}$ and the CERN $e^+e^-$ (LEP) Colliders~\cite{lep}.  The CDF collaboration reported evidence of $ZZ$ production with a significance of 4.4 
standard deviations from combined $ZZ\to\ell^{+}\ell^{-}\ell^{'+}\ell^{'-}$ 
and $ZZ\to\ell^{+}\ell^{-}\nu\bar{\nu}$ searches and  measured a production cross 
section of $\sigma(ZZ)=1.4^{+0.7}_{-0.6}$~pb with $1.9$~fb$^{-1}$ of integrated luminosity~\cite{cdf_zz}.   The D0 collaboration reported an observation of $ZZ\to\ell^{+}\ell^{-}\ell^{'+}\ell^{'-}$ 
($\ell$, $\ell^{'}$ = $e$ or $\mu$)
with 1.7 fb$^{-1}$ of data and measured the production cross section to be $\sigma(ZZ) = 1.75^{+1.27}_{-0.86}~\mathrm{(stat)} \pm 
0.13~\mathrm{(syst)}$~pb~\cite{runiib_zz}.  That result was combined with a previous $ZZ\to4\ell$ analysis~\cite{runiia_zz} and an analysis in the $ZZ\to\ell^{+}\ell^{-}\nu\bar{\nu}$ channel~\cite{d0zzllnn}, giving a cross section of $\sigma(ZZ) = 1.60\pm0.63~\mathrm{(stat)}^{+0.16}_{-0.17}~\mathrm{(syst)}$~pb with a significance of 5.7 standard deviations~\cite{runiib_zz}.

In this Article, we present a measurement of $Z/\gamma^*$ boson pair production with subsequent decays to either
electron or muon pairs, resulting in final states consisting
of four electrons ($4e$), four muons ($4\mu$), or two muons and
two electrons ($2\mu2e$)~\cite{particles}.  We accept events which have more than four leptons, however we only use the four leptons with highest transverse momenta in constructing kinematic variables.  Compared to previous publications~\cite{cdf_zz,runiib_zz} we use a larger dataset and more inclusive selection criteria to achieve a reduction of a factor of 2.5 for the statistical uncertainty which dominates the experimental cross section determination.  The higher statistics opens the possibility of studies of $Z/\gamma^*Z/\gamma^*$ production properties, and we present for the first time differential distributions for the final state particles.  Data used in this analysis were collected with the D0~detector at the Fermilab Tevatron $p\bar{p}$ Collider at $\sqrt{s}=1.96$ TeV between April 2002 and March 2010 and correspond to an integrated luminosity of $6.4 \pm 0.4$~fb$^{-1}$~\cite{d0lumi}.  


The D0 detector~\cite{run2det} consists of a central tracking system, a calorimeter, and a muon detection system.  A silicon microstrip tracker (SMT) and a scintillating fiber tracker (CFT) comprise the tracking system, which provides coverage for pseudorapidity
$|\eta_{\text{det}}| < 3$ ~\cite{pseudo}.  The tracking systems are located within a 2~T superconducting solenoidal magnet.  Located immediately before the inner layer of the calorimeter is the central preshower detector (CPS), consisting of approximately one radiation length of absorber followed by three layers of scintillating strips.  Calorimetry is provided by three liquid argon and uranium calorimeters.  The central calorimeter (CC) provides coverage for $|\eta_{\text{det}}| < 1.1$, while the two end-cap calorimeters (EC) extend coverage to $|\eta_{\text{det}}| < 3.2$.  The calorimeters are sectioned in order of increasing distance from the collision point.  The section closest to the collision region is the electromagnetic section (EM), while farther away are the fine hadronic (FH), and the coarse hadronic (CH) sections.  A muon system surrounds the calorimeters, consisting of three layers of scintillators and drift tubes and 1.8~T iron toroidal magnets, covering $|\eta_{\text{det}}| < 2$.

All events used in this analysis are recorded after satisfying a mixture of single and dilepton triggers.  Due to the high transverse momentum of the $Z/\gamma^*$ decay products and the number of leptons in the final state, the trigger efficiency exceeds 99\%. 

The $4e$ channel requires the presence of four electrons with transverse energies $E_{T} >$ 30, 25, 15, and 15~GeV, respectively. 
Electrons can be reconstructed in either the CC region or in the
EC region, however at least two electrons must be in the CC region.  Electrons must be isolated from other energy clusters in the calorimeter and have a large fraction of their energy deposited in the EM section of the calorimeter.  Electrons in the CC are required to satisfy identification criteria 
based on multivariate discriminants which use calorimeter shower shape, CPS, and tracking information.  Several of these parameters are inputs to a neural network (NN), which is used to enhance electron purity.  Electrons in the CC are required to have a matched track in the central tracking system.  Electrons in the EC are not required to have a track matched to them due to deteriorating tracking coverage for $|\eta_{\text{det}}| > 2$, but must satisfy additional shower shape requirements as well as pass tighter NN selections.  With no requirement applied on the charge of the electrons to increase selection efficiency,
three possible $Z/\gamma^*Z/\gamma^*$ combinations can be formed for each $4e$
event.  Only events having an invariant mass pair $>70$~GeV and the other pair $>50$~GeV are considered.  Finally, events are split
into three categories, depending on the number of
electrons in the CC region.  Subsamples with two, three, and four electrons in the CC are denoted as $4e_{2C}$, $4e_{3C}$,
and $4e_{4C}$, respectively.  This splitting is performed because these subsamples have different levels of background contamination. 

For the $4\mu$ channel, muons are identified as track segments in the muon detector matched to a central track or as a central track matched to a pattern of calorimeter activity consistent with passage of a high momentum muon.  Muons identified in the muon system must satisfy quality criteria based on scintillator and wire information, and be synchronous with the beam crossing time to reject background from cosmic rays.  At least three muons in the event must be isolated.  Muon isolation is dependent upon two cone-based variables.  The first variable, $T_{\text{Halo}}$, is the sum of the transverse momentum associated with tracks in a cone of radius ${\Delta\cal R}=\sqrt{(\Delta\eta)^{2} + (\Delta\phi)^{2}} = 0.4$ centered on the muon track.  The second variable, $C_{\text{Halo}}$, is the transverse energy measured in the calorimeter, in an annulus between ${\Delta\cal R}=0.1$ and ${\Delta\cal R}=0.4$ centered on the muon track.  Muons with muon system reconstructed tracks are considered isolated if $T_{\text{Halo}}$ is less than $4$ GeV.  For muons with only a calorimeter signal or where the muon system provides track segments only, a tighter isolation requirement is used: $T_{\text{Halo}}/p_{T}^{\mu} < 0.09$ and $(C_{\text{Halo}}-0.005 {\cal L})/p_{T}^{\mu} < 0.09$, where $p_{T}^{\mu}$ is the transverse momentum of the muon track, and ${\cal L}$~represents the instantaneous luminosity (in units of $10^{30}$~cm$^{-2}$s$^{-1}$, ${\cal L}$ can reach $\approx$300) which is introduced to account for the occupancy increase due to multiple $p\bar{p}$ interactions at higher luminosities.  We require that the four most energetic muons have ordered transverse momenta $p_T >$ 30, 25, 15, and 15~GeV, respectively.  The difference between the distances of closest approach ($dca$) to the $p\bar{p}$ interaction point along the beam axis for any pair of muon tracks are required to be $<$ 3.0 cm.  The three possible $Z/\gamma^*Z/\gamma^*$ combinations per event formed without considering muon charge are considered.  Candidate events are selected when at least one of the three possible combinations satisfies the same dilepton invariant mass requirements applied in the $4e$ channel.


For the $2\mu2e$ channel, one electron and one muon must have $E_T(p_T)>20$ GeV, while the other two leptons must have $E_T(p_T)>15$ GeV.  All muons and electrons must satisfy the lepton selection criteria 
defined for the $4e$ and $4\mu$ final states, except that only one muon must satisfy the isolation requirements 
imposed in the $4\mu$ final state.  In addition, electrons and muons are
required to be spatially 
separated by ${\Delta\cal R} > 0.2$.  This requirement is applied to remove $Z \to \mu\mu$ background
where the muons radiate photons leading to events with two muons and two 
trackless electron candidates.  Events from this channel assume that the muon pair originated from one $Z/\gamma^*$ and the electron pair originated for the other $Z/\gamma^*$.  The two same-flavor lepton pairs are required to satisfy the same invariant mass requirements as for the $4e$ channel.  Finally, events are split
into three categories depending on the number of
electrons in the CC region. Subsamples with zero, one, and two or more electrons in the CC are denoted
as $2\mu2e_{0C}$, $2\mu2e_{1C}$, and 
$2\mu2e_{2C}$, respectively.  As in the $4e$ channel, this splitting is performed because these subsamples have different levels of background contamination.

A Monte Carlo (MC) simulation is used to determine signal acceptances, efficiencies as well as 
the expected number of signal events in each subchannel.  All signal acceptances and efficiencies are evaluated after the high mass ($>70$~GeV and $>50$~GeV) requirements have been applied at the MC generator level.  The contribution from $Z/\gamma^*Z/\gamma^*$ events with at least one $Z/\gamma^*$ boson decaying into tau pairs is included in the
signal.  Events are generated using \textsc{pythia} and passed through
a detailed \textsc{geant}-based~\cite{geant} simulation of the detector
response.  Differences between MC and data reconstruction and identification efficiencies for
electrons and muons are corrected using efficiencies derived from 
large data samples of inclusive $Z\rightarrow \ell\ell$ events.

Backgrounds to the $Z/\gamma^*Z/\gamma^*$ signal originate from events with $W$ and/or $Z$ bosons decaying to leptons 
plus additional jets or photons and from top quark pair ($t\bar{t}$)
production.  The jets can be misidentified as leptons or contain electrons or muons from in-flight decays of pions,
kaons, or heavy-flavored hadrons. 

To estimate the background from events with misidentified leptons,
we first measure the probability for a jet to produce an 
electron or muon that satisfies the identification criteria from data.  We measure this probability in a separate dijet data sample, selected by requiring at least two jets with $p_T > 15$~GeV.  We require the jet with largest $p_T$ to pass strict jet identification criteria and we use the second jet to measure the probability for a jet to be misidentified as a lepton.  The two jets are required to be separated in azimuth by $\Delta \phi > 3.0$.  To suppress contamination from $W+$jet events, we require the missing transverse energy \Etmiss~$<$ 20 GeV~\cite{met}.  The lepton identification criteria are applied to the second jet to measure how often a jet mimics an electron or produced a muon.


The probability for a jet to mimic an electron, parameterized in jet $E_T$ and $\eta$, is approximately $4 \times 10^{-4}$ 
for the case of CC electrons with a 
matched track and approximately $2 \times 10^{-3}$
in the case of EC electrons for which no track match criterion is applied. 
The probabilities for jets to be misidentified as
electrons are then applied to jets in $eee$+jets and
$\mu\mu e$+jets data to determine the background to the 
$4e$ and $2\mu2e$
channels, respectively. This method takes into account
contributions from $Z$+jets, $Z$+$\gamma$+jets,
$WZ$+jets, $WW$+jets, $W$+jets, and events with $\ge4$ jets. 
However, it counts the contribution from  
$Z$+jets twice. A correction is measured using data, 
amounting to approximately $10$\%.  

The probability for a 15~GeV (100~GeV) jet to produce a muon of 
$p_T>15$ GeV is approximately $7 \times 10^{-4} \ (10^{-2}) $ without requiring muon isolation,
and approximately $4 \times 10^{-4} \ (2 \times 10^{-3})$
when the muon is required to be isolated.  The probabilities for jets to contain a muon are applied to
jets in $\mu\mu$+jets and $ee$+jets data to estimate the background
for the $4\mu$ and $2\mu2e$ channels.
		
The background from $t\bar{t}$ production is estimated from simulation with \textsc{alpgen}~\cite{alpgen} generated events interfaced to \textsc{pythia}~\cite{pythia}.  


Another possible background in the $4\mu$ and $2\mu2e$ channels is from cosmic ray muons.  The probability for cosmic ray muons to cross at the interaction region near the time of the $p\bar{p}$ collision is small, nonetheless we estimate this background using data.  The estimation is done by reversing combinations of the $4\mu$ sample selection requirements, such as scintillator timing and $dca$ criteria.  This procedure yields rejection factors which are then applied to a cosmic ray enhanced data sample.  The resulting background from cosmic rays in the $4\mu$ and $2\mu2e$ samples is less than $0.01$ event for each channel.  

We also estimate the contribution of $Z/\gamma^*Z/\gamma^*$ with low invariant mass lepton pairs ($<70$~GeV and $<50$~GeV) that pass the kinematic selection criteria due to detector and reconstruction effects.  This migration contribution is found from our signal MC where we select events that fail the generator level mass selection.  This small contribution is corrected for in the cross section measurement.  

\begin{table*}[hbt]
\caption{The expected number of $Z/\gamma^*Z/\gamma^*$ and 
background events [$t\bar{t}$, $W$/$Z$/$\gamma$+jets, and cosmic ray
contributions], and the number of observed candidates in the 
seven $Z/\gamma^*Z/\gamma^* \to\ell^{+}\ell^{-}\ell^{'+}\ell^{'-}$ subchannels.  The expected number of $Z/\gamma^*Z/\gamma^*$ events assumes the NLO theoretical cross section of $1.4$ pb.  Uncertainties
reflect statistical and systematic contributions added in quadrature.}
\vspace*{2mm}
\label{tab:channels}
\begin{tabular}{cccccccc} 
\hline \hline
Subchannel 
& $4e_{2C}$ 
& $4e_{3C}$ 
& $4e_{4C}$ 
& $4\mu$             
& $2\mu2e_{0C}$ 
& $2\mu2e_{1C}$ 
& $2\mu2e_{2C}$ \\ 
\hline
 & & & & & & & \\ [-3mm]
$Z/\gamma^*Z/\gamma^*$ 
& $0.31 \pm 0.05$ 
& $0.73 \pm 0.12$
& $0.69 \pm 0.11$ 
& $2.57 \pm 0.36$            
& $0.24 \pm 0.03$ 
& $1.41 \pm 0.18$ 
& $2.58 \pm 0.33$ \\ 
& & & & & & & \\ [-2mm]
$Z/\gamma^*Z/\gamma^*$ Migration
& $0.019^{+0.007}_{-0.004}$ 
& $0.027^{+0.006}_{-0.005}$ 
& $0.020^{+0.008}_{-0.006}$ 
& $0.106^{+0.027}_{-0.015}$ 
& $0.002^{+0.002}_{-0.001}$ 
& $0.002^{+0.001}_{-0.001}$ 
& $0.008^{+0.003}_{-0.002}$ \\ 
 & & & & & & & \\ [-2mm]
$W$/$Z$/$\gamma$+jets
& $0.065 \pm 0.013$ 
& $0.041 \pm 0.007$ 
& $0.024 \pm 0.007$ 
& $0.035 \pm 0.015$ 
& $0.030^{+0.011}_{-0.009}$ 
& $0.057^{+0.010}_{-0.009}$ 
& $0.078^{+0.015}_{-0.014}$ \\[-2mm] 
 & & & & & & & \\
Cosmics
& $\dotsm$ 
& $\dotsm$ 
& $\dotsm$ 
& $<$ 0.01 
& $<$ 0.001 
& $<$ 0.003 
& $<$ 0.006       \\
 & & & & & & & \\ [-3mm]
$t\bar{t}$ 
& $\dotsm$ 
& $\dotsm$ 
& $\dotsm$
& $\dotsm$ 
& $0.0013^{+0.0010}_{-0.0009}$ 
& $0.0138^{+0.0070}_{-0.0069}$ 
& $0.0091^{+0.0041}_{-0.0039}$ \\[1mm] 

Observed events       
& 0       
& 1        
& 2          
& 4             
& 0       
& 1        
& 2 \\ 
\hline \hline
\end{tabular}
\end{table*}

Table~\ref{tab:channels} summarizes the expected signal and background
contributions to each subchannel, as well as the numbers of candidate
events in data.  The systematic uncertainty for the signal yield
is dominated by a 6\% uncertainty on the luminosity measurement~\cite{d0lumi}, the theoretical cross section uncertainty of 7\%, and the uncertainty on the four-lepton reconstruction efficiencies of $\approx10$\%.  Additional smaller systematic uncertainties arise from modeling energy and momentum resolutions and from MC modeling of the signal kinematics.  A systematic uncertainty of 20\% on the jet-to-electron misidentification
probability is estimated by varying the selection criteria of the control samples.  Systematic uncertainties on background from jets containing a muon arise from the 40\% uncertainty in measured misidentification rates and from the limited statistics of the data remaining in the samples
after selection. The $t\bar{t}$ background systematic uncertainty includes the 7\% uncertainty on $\sigma(t\bar{t})$, 
as well as contributions from the variation in cross section
and acceptance originating from the uncertainty on the mass of
the top quark. 

\begin{figure}[h!]
	\includegraphics[scale=0.35,angle=0]{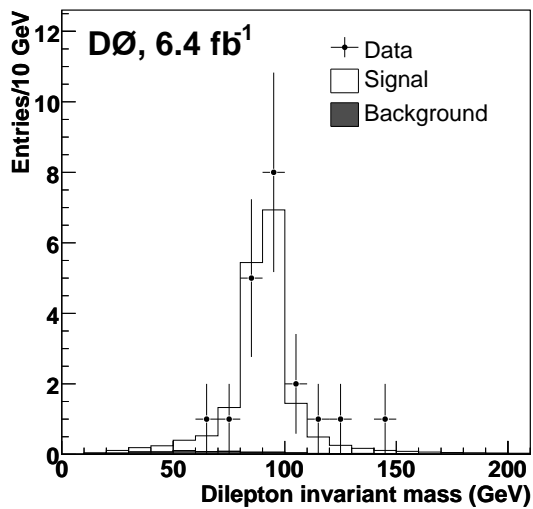}
		\caption{\label{fig:dileptmass} Distribution of the dilepton masses compared to the expected signal and background.}
\end{figure}

\begin{figure}[h!]

    \includegraphics[scale=0.35,angle=0]{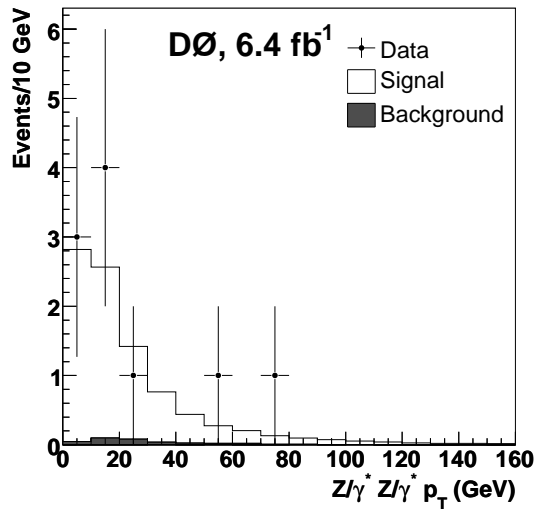}
     
    \caption{Distribution of $Z/\gamma^*Z/\gamma^*$ $p_{T}$ compared to the expected signal and background.}
    \label{fig:zzpt}
  
\end{figure}

\begin{figure}[h!]

    \includegraphics[scale=0.35,angle=0]{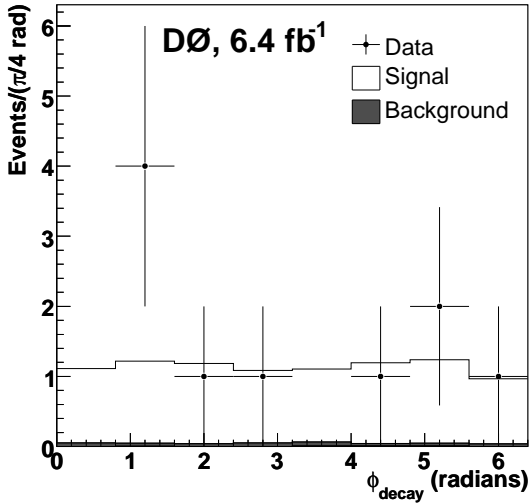}
     \caption{Distribution of the azimuthal angle $\phi_{\text{decay}}$ for the decay planes of the $Z/\gamma^*$ bosons compared to the expected signal and background.}
    \label{fig:zzdelphi}
  
\end{figure}

\begin{figure}[h!]
	\includegraphics[scale=0.35,angle=0]{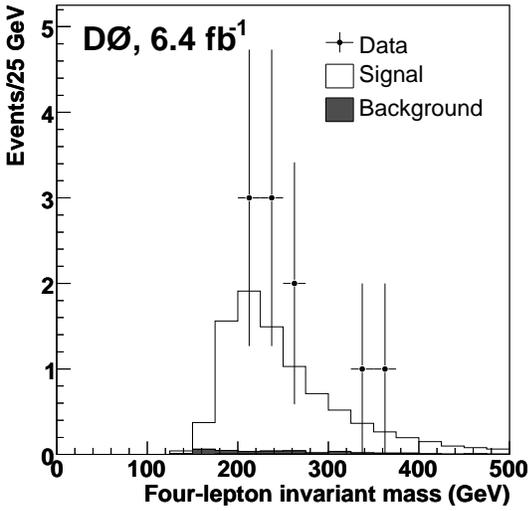}
	\caption{\label{fig:fourmass} Distribution of four-lepton invariant mass compared to the expected signal and background.}
\end{figure}

The expected number of signal and background events are $8.73 \pm 1.22$ and $0.37 \pm 0.13$, respectively.  We observe a total of ten candidate events, three
in the $4e$ subchannel, four in the $4\mu$
subchannel, and three in the $2\mu 2e$ subchannel.  


Figures~\ref{fig:dileptmass}--\ref{fig:fourmass} show four kinematic distributions of the data compared to the expected signal and background.  In the $eeee$ and $\mu\mu\mu\mu$ subchannels there can be up to three
possible pairings of the four leptons which satisfy the invariant mass requirements
used to select candidate events.  If two or more combinations satisfy the invariant mass requirements we select the one in which both dilepton pairs
have an invariant mass closest to the nominal $Z$ boson mass for the distributions shown in Figs.~\ref{fig:dileptmass} and ~\ref{fig:zzdelphi}.  Figure~\ref{fig:dileptmass} shows the distribution of dilepton masses (two entries per event), Fig.~\ref{fig:zzpt} the transverse momentum of the $Z/\gamma^*Z/\gamma^*$ system.  Figure~\ref{fig:zzdelphi} displays the azimuthal angle $\phi_{\text{decay}}$, i.e. the angle through which the lepton side of one of the $Z/\gamma^*$ boson decay planes is rotated into the lepton side of the other $Z/\gamma^*$ boson decay plane, as measured in the $Z/\gamma^*Z/\gamma^*$ center-of-mass frame.  This angle is discriminating against background for high mass Higgs bosons.  The construction of $\phi_{\text{decay}}$ used in this Article follows the definition in~\cite{ZZdelphi}.  Figure~\ref{fig:fourmass} displays the invariant mass of the $Z/\gamma^*Z/\gamma^*$ system.  Additional differential distributions and event information for the selected sample of events are shown in~\cite{supple}.    

The distributions shown are consistent with the expectation of a $Z/\gamma^*Z/\gamma^*$ signal and small background.  We therefore proceed to measure the $p\bar{p}\rightarrow Z/\gamma^*Z/\gamma^*$ production cross section $\sigma$.  Using the following likelihood function:

\vspace{-5mm}
\begin{eqnarray}
L(N_j^{\text{obs}}, \mu_j )={\prod_{j=1}^{7}}\frac{\mu_j^{N^{\text{obs}}}}{N_j^{\text{obs}}!}e^{-\mu_j} ,
\end{eqnarray}
\vspace{-4mm}

\noindent where $N_j^{\text{obs}}$ is the observed number of events given an expected signal and 
background yield of

\vspace{-5mm}
\begin{eqnarray}
\mu_j=\sigma\times A_j \times {\cal{B}}_j\times{\cal{L}}_j+N_j^{\text{bkgd}}.
\end{eqnarray}
\vspace{-5mm}

\noindent Here, $A_j$ is the acceptance times efficiency, ${\cal{L}}_j$ is 
the integrated luminosity,  ${\cal{B}}_j$ is the branching fraction, and
$N_j^{\text{bkgd}}$ is the expected background for subchannel $j$.  The cross section $\sigma$ is obtained by minimizing 
$-\ln(L)$.
The statistical uncertainty on $\sigma$ is obtained by varying the $-\ln(L)$ by half a unit above the minimum.  Systematic uncertainties are propagated to cross section uncertainties via variations in the likelihood function due to each independent systematic source.  These likelihood variations are then summed in quadrature to obtain the total systematic uncertainty.  

The production cross section is measured to be $\sigma(p\bar{p} \to Z/\gamma^*Z/\gamma^*)=1.33^{+0.50}_{-0.40} 
~\mathrm{(stat)} \pm 0.12 ~\mathrm{(syst)} \pm 0.09~\mathrm{(lumi)}$~pb.  This result is consistent with the SM prediction of $1.4\pm0.1$ pb.  The total uncertainty reflects an improvement by a factor of approximately 2.5 relative to our previous four charged lepton measurement~\cite{runiib_zz}.  Based on this result we also quote a measurement of the on-shell $\sigma(p\bar{p} \to ZZ)$ cross section.  Using the conversion factor of 0.93 found from simulation, we measure $\sigma(p\bar{p} \to ZZ)=1.24^{+0.47}_{-0.37} ~\mathrm{(stat)} \pm 0.11 ~\mathrm{(syst)} \pm 0.08~\mathrm{(lumi)}$~pb.  


The significance of the observed event distribution
is found by using a negative log-likelihood ratio (NLLR) 
test statistic defined as $-2\ln ({L_{S+B}}/{L_{B}} )$, where $L_{B}$ and $L_{S+B}$ are Poisson likelihood functions for background and signal plus background, respectively~\cite{collie}. As input we use the expected numbers of events from signal and background, separated into
the seven subchannels, compared to the observed numbers of data events.  The significance is obtained by generating many pseudo-experiments which are created by varying the signal and background around their central predicted values, thus creating a distribution of NLLRs.  The mean numbers of expected signal and background events per pseudo-experiment are varied according to their systematic uncertainties.  The method gives the probability ($p$-value) of the background fluctuating
to give the observed yields or higher. In 2$\times10^9$
background pseudo-experiments, we find zero trials with an NLLR value
smaller or equal to that observed in data. This gives
a $p$-value of less than 10$^{-9}$.  The equivalent probability for a Gaussian distribution is greater than 6 standard deviations.  

Finally, this result is combined with the result from the independent $ZZ \to \ell^{+}\ell^{-}\nu\bar{\nu}$ analysis~\cite{d0zzllnn}.  The combination is done by adding the $ZZ \to \ell^{+}\ell^{-}\nu\bar{\nu}$ results in dielectron and dimuon final states to our likelihood calculation as additional channels.  Correlations of systematic uncertainties are accounted for between the two analyses.  The combined result is $\sigma(p\bar{p} \to ZZ)=1.40^{+0.43}_{-0.37} ~\mathrm{(stat)} \pm 0.14 ~\mathrm{(syst)}$~pb.    

In summary, the $Z/\gamma^*Z/\gamma^*$ cross section in $p\bar{p}$ interactions at $\sqrt{s}$=1.96 TeV is measured to be $1.33^{+0.50}_{-0.40}~\mathrm{(stat)} \pm 0.12 ~\mathrm{(syst)} \pm 0.09~\mathrm{(lumi)}$~pb.  The on-shell $ZZ$ production cross section is $1.24^{+0.47}_{-0.37} ~\mathrm{(stat)} \pm 0.11 ~\mathrm{(syst)} \pm 0.08~\mathrm{(lumi)}$~pb.  The new D0 combined result is $\sigma(p\bar{p} \to ZZ)=1.40^{+0.43}_{-0.37} ~\mathrm{(stat)} \pm 0.14 ~\mathrm{(syst)}$~pb.  These results constitute the most precise measurement to date of the $p\bar{p} \to Z/\gamma^*Z/\gamma^*$ and $p\bar{p} \to ZZ$ cross sections and demonstrate sufficient statistics for an examination of $Z/\gamma^*Z/\gamma^*$ kinematic distributions.  The kinematic distributions of the 10 observed events are consistent with the SM predictions.

%
We thank the staffs at Fermilab and collaborating institutions,
and acknowledge support from the
DOE and NSF (USA);
CEA and CNRS/IN2P3 (France);
FASI, Rosatom and RFBR (Russia);
CNPq, FAPERJ, FAPESP and FUNDUNESP (Brazil);
DAE and DST (India);
Colciencias (Colombia);
CONACyT (Mexico);
KRF and KOSEF (Korea);
CONICET and UBACyT (Argentina);
FOM (The Netherlands);
STFC and the Royal Society (United Kingdom);
MSMT and GACR (Czech Republic);
CRC Program and NSERC (Canada);
BMBF and DFG (Germany);
SFI (Ireland);
The Swedish Research Council (Sweden);
and
CAS and CNSF (China).
%

\newpage
\begin{widetext}
\section{\label{app:supplement} Supplemental Material}

Figures~\ref{fig:PT}--\ref{fig:dimass} show various kinematic quantities for data and for the expected signal and background.
Figures~\ref{fig:PT} and~\ref{fig:deteta} show individual lepton $p_{T}$ and $\eta_{\text{det}}$, where the leptons are ranked from highest transverse momentum to lowest~\cite{pseudo}.  Figure~\ref{fig:delR} shows $\Delta \phi$ and $\Delta \cal{R}$ between two leptons (two entries per event).  Figure~\ref{fig:zpt} shows the dilepton pair $p_{T}$ ranked from highest transverse momentum to lowest.  In the $eeee$ and $\mu\mu\mu\mu$ subchannels there can be up to three possible pairings of the four leptons which satisfy the invariant mass requirements
used to select candidate events.  If two or more combinations satisfy the invariant mass requirements we select the one in which both dilepton pairs
have an invariant mass closest to the nominal $Z$ boson mass.  For the $2\mu2e$ subchannel, we use the only valid combination, where one $Z/\gamma^*$ decays to the $ee$ pair and the other $Z/\gamma^*$ decays to the $\mu\mu$ pair.  Figure~\ref{fig:dimass} shows the distribution of dilepton mass $M_2$ versus dilepton mass $M_1$, where $M_1$ is the dilepton mass associated with the lepton pair that has the higher $p_T$.  Figure~\ref{fig:plane_ZZ} gives an illustrative description of the angle $\phi_{\text{decay}}$.  Tables~\ref{tab:emcand1}--\ref{tab:emucand3} give various measured quantities of the ten candidate events.

\begin{figure}[h!]
	\includegraphics[scale=0.50,angle=0]{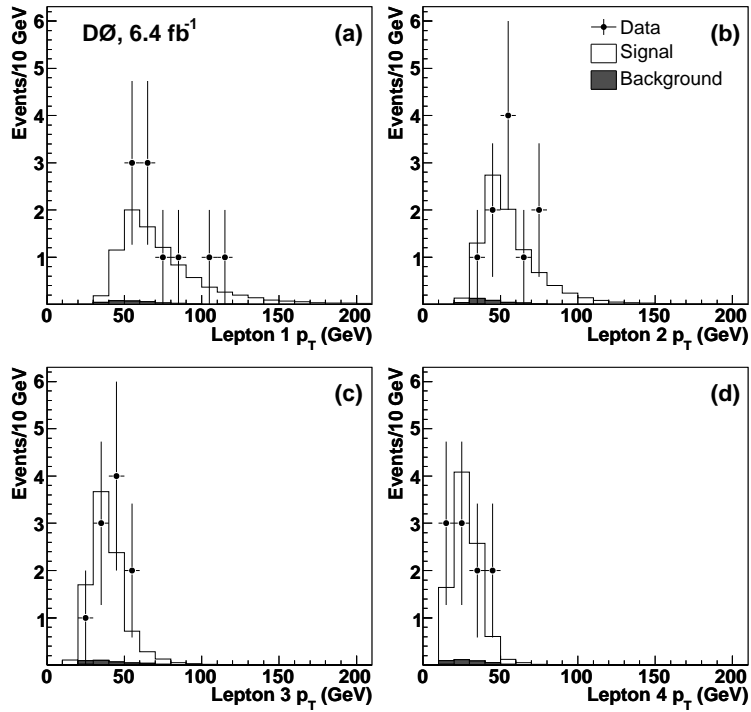}
	
		\caption{\label{fig:PT} Distributions of transverse momentum compared to the expected signal and background for the (a) leading, (b) second, (c) third, and (d) fourth leptons.}
\end{figure}

\begin{figure}[h!]

    \includegraphics[scale=0.50,angle=0]{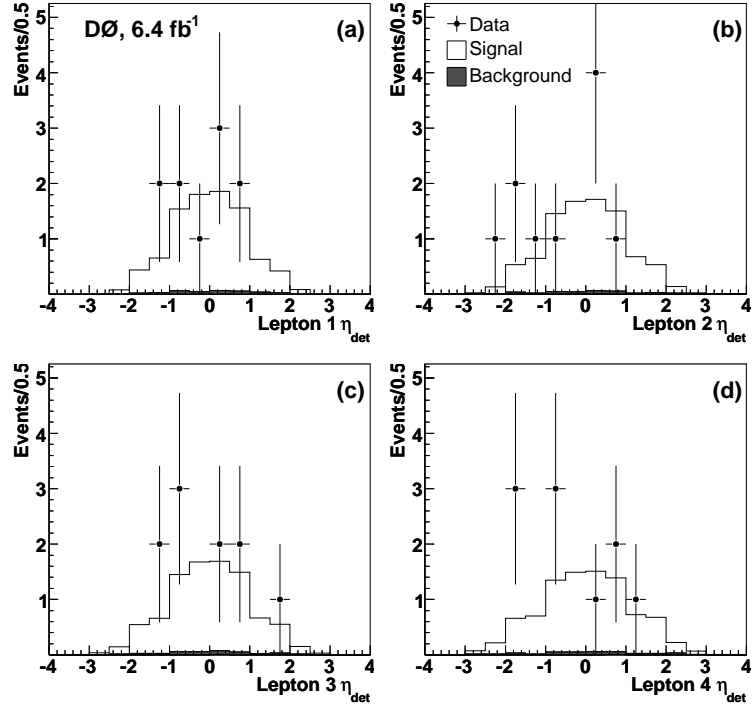} 
   \caption{Distributions of $\eta_{\text{det}}$ compared to the expected signal and background for the (a) leading, (b) second, (c) third, and (d) fourth leptons.}
    \label{fig:deteta}
  
\end{figure}

\begin{figure}[h!]
		\includegraphics[scale=0.3,angle=0]{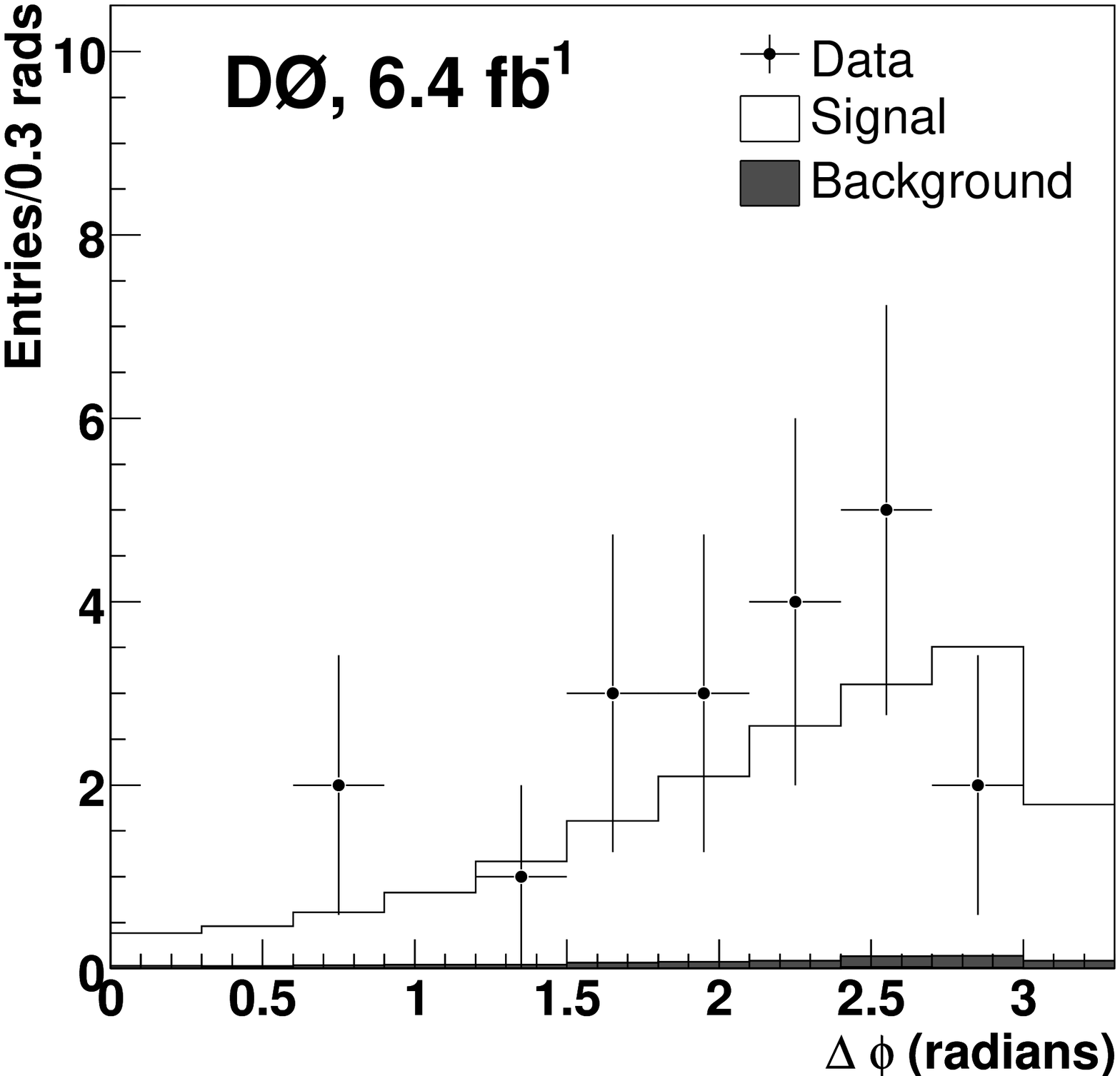}
  \includegraphics[scale=0.3,angle=0]{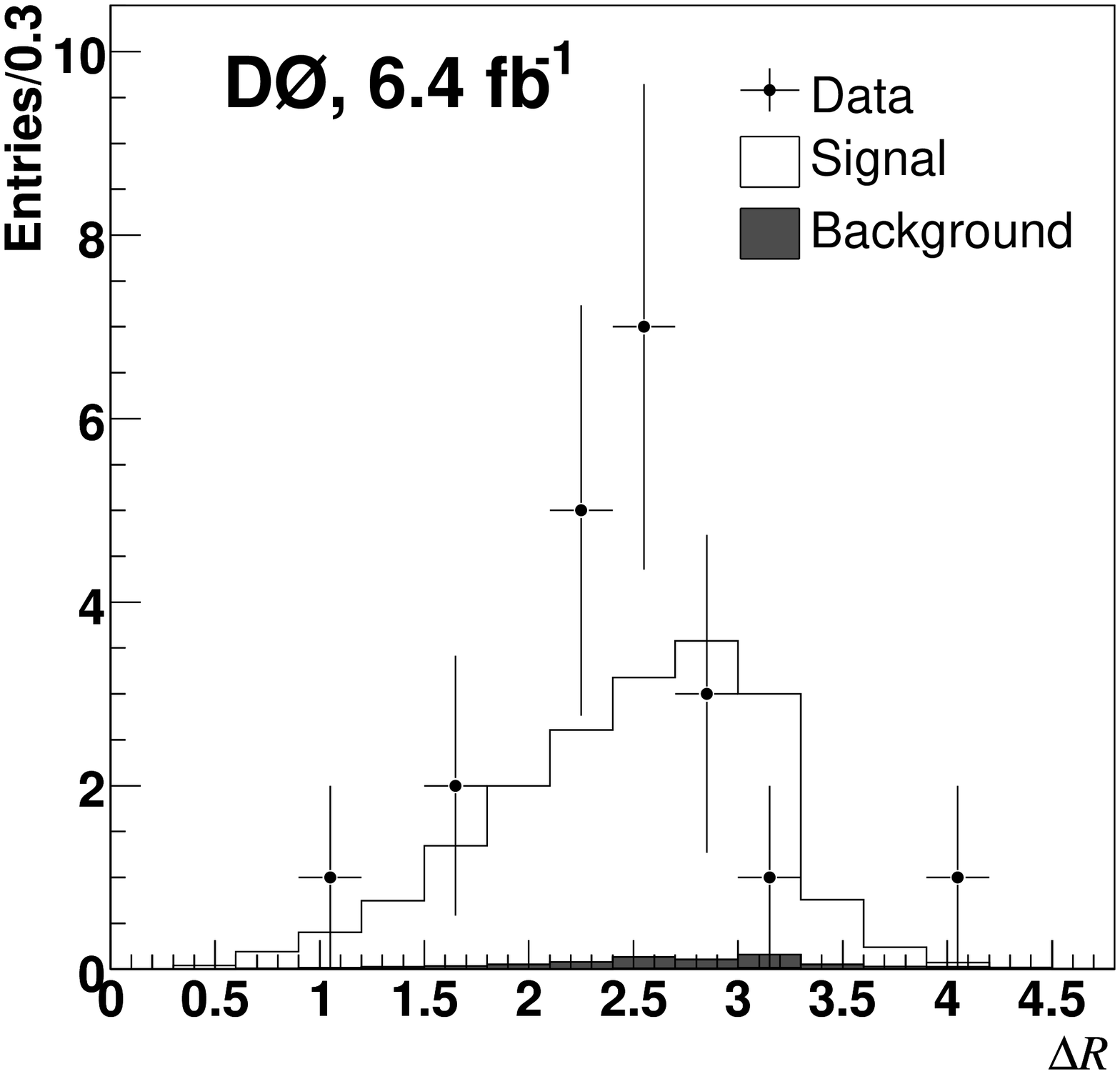}
   \caption{Distributions of $\Delta \phi$ and $\Delta \cal{R}$ between leptons compared to the expected signal and background.  For the eeee and $\mu\mu\mu\mu$ channels, the combinations shown are those which are most consistent with a $Z$ boson mass hypothesis of 91.2 GeV.}
    \label{fig:delR} 
\end{figure}

\begin{figure}[h!]
    \includegraphics[scale=0.6,angle=0]{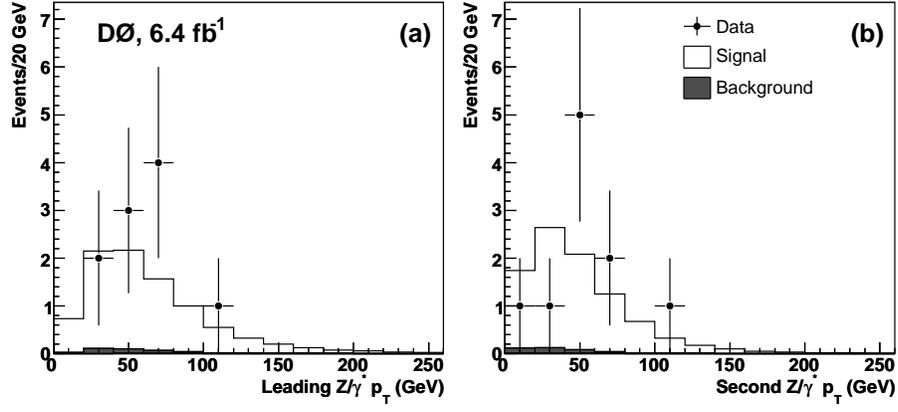}
    \caption{Distributions of $Z/\gamma^*$ $p_T$ for lepton pairings with
    (a) highest and (b) lowest  $Z/\gamma^*$ $p_T$ compared to the expected signal and background.}
    \label{fig:zpt}
\end{figure}

\begin{figure}[h!]
	\includegraphics[scale=0.4,angle=0]{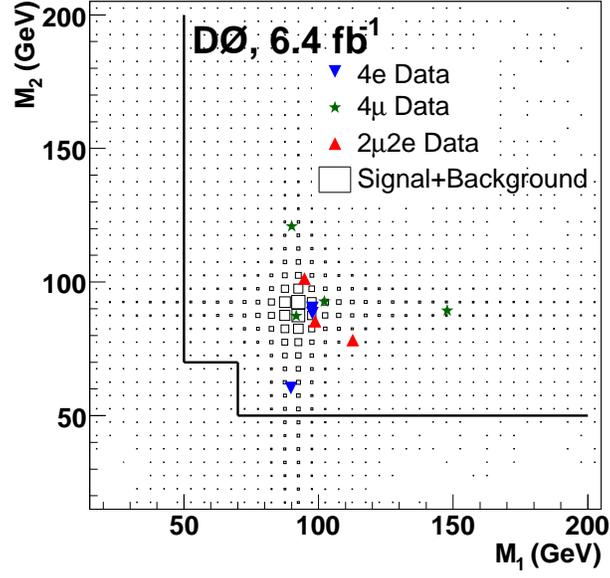}
				\caption{\label{fig:dimass} Dilepton mass pairing compared to the expected signal and background.  The lines indicate where the invariant mass requirements are applied.}
\end{figure}

\begin{figure}[h!]

    \includegraphics[scale=0.8,angle=0]{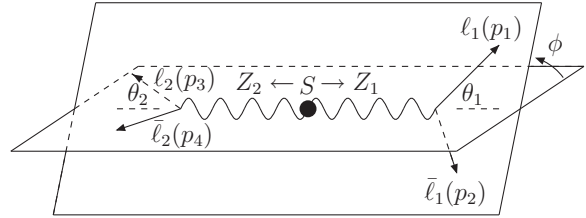}
     
    \caption{The definition of the azimuthal angle $\phi_{\text{decay}} 
   \equiv \phi \in [0,2\pi]$.  $\bar{l}_1$ and $\bar{l}_2$ represent the anti-leptons from $Z$ bosons 1 and 2, respectively.  Obtained from~\cite{ZZdelphi}.}
    \label{fig:plane_ZZ}
  
\end{figure}

\begin{table}[h!]
  \caption{\label{tab:emcand1}Summary of the properties of the 1st $eeee$ candidate event.}
  \begin{tabular}{lcccc} 
    \hline \hline
    Run Number & \multicolumn{4}{c}{231347} \\ 
    Event Number & \multicolumn{4}{c}{25076242} \\
    \hline
    & $e_1$ & $e_2$ & $e_3$ & $e_4$ \\
    \hline
    $p_T$ (GeV) & $107 \pm 4$ & $59\pm3$ & $52\pm2$ & $17\pm1$ \\
    
    $\eta$ & 0.66 & 0.25 & -0.64 & -0.85 \\
    
    $\phi$ (radians) & 4.09 & 1.08 & 0.46 & 2.62 \\
    
    $z_{\text{vtx}}$ (cm) & 3.4 & 3.4 & 3.4 & 3.4 \\
    
    Charge & +1 & +1 & $-1$ & $-1$ \\
    \hline
    &       &        &        & \\ [-3mm]
    & \multicolumn{2}{c}{$e_{1}^{+}e_{4}^{-}$} & 
    \multicolumn{2}{c}{$e_{2}^{+}e_{3}^{-}$} \\ [1mm]
    \hline
    Dilepton mass (GeV) & \multicolumn{1}{r}{89} & \multicolumn{1}{l}{$\pm$ 3} & \multicolumn{1}{r}{60} & \multicolumn{1}{l}{$\pm$ 2} \\
    $Z/\gamma^* p_{T}$ (GeV) & \multicolumn{1}{r}{110} & \multicolumn{1}{l}{$\pm$ 4} & \multicolumn{1}{r}{106} & \multicolumn{1}{l}{$\pm$ 3} \\
    \hline
    4-lepton mass (GeV) & \multicolumn{2}{r}{274} & \multicolumn{2}{l}{$\pm$ 6} \\
    $Z/\gamma^*Z/\gamma^* p_{T}$ (GeV) & \multicolumn{2}{r}{5} & \multicolumn{2}{l}{$\pm$ 5} \\
    \Etmiss \ (GeV) & \multicolumn{2}{r}{3} & \multicolumn{2}{l}{$\pm$ 2} \\
    \hline \hline
  \end{tabular}
\end{table}

\begin{table}
  \caption{\label{tab:emcand2} Summary of the properties of the 2nd $eeee$ candidate event.}
  \begin{tabular}{lcccc} 
    \hline \hline
    Run Number & \multicolumn{4}{c}{223736} \\ 
    Event Number & \multicolumn{4}{c}{14448774} \\
    \hline
    & $e_1$ & $e_2$ & $e_3$ & $e_4$ \\
    $p_T$ (GeV) & $83\pm3$ & $75\pm3$ & $35\pm2$ & $27\pm2$ \\
    $\eta$ & 0.64 & 0.39 & 0.85 & 1.18 \\
    $\phi$ (radians) & 6.16 & 3.80 & 3.83 & 1.40 \\
    $z_{\text{vtx}}$ (cm) & -19.2 & -19.2 & -19.2 & -19.2 \\
    Charge & +1 & +1 & $-1$ & $-1$ \\
    \hline
    &       &        &        & \\ [-3mm]
    & \multicolumn{2}{c}{$e_{1}^{+}e_{3}^{-}$} & 
    \multicolumn{2}{c}{$e_{2}^{+}e_{4}^{-}$} \\ [1mm]
    \hline
    Dilepton mass (GeV) & \multicolumn{1}{r}{98} & \multicolumn{1}{l}{$\pm$ 3} & \multicolumn{1}{r}{88} & \multicolumn{1}{l}{$\pm$ 4} \\
    $Z/\gamma^* p_{T}$ (GeV) & \multicolumn{1}{r}{63} & \multicolumn{1}{l}{$\pm$ 3} & \multicolumn{1}{r}{57} & \multicolumn{1}{l}{$\pm$ 3} \\
    \hline
    4-lepton mass (GeV) & \multicolumn{2}{r}{216} & \multicolumn{2}{l}{$\pm$ 5} \\
    $Z/\gamma^*Z/\gamma^* p_{T}$ (GeV) & \multicolumn{2}{r}{51} & \multicolumn{2}{l}{$\pm$ 3} \\
    \Etmiss \ (GeV) & \multicolumn{2}{r}{18} & \multicolumn{2}{l}{$\pm$ 13} \\
    \hline \hline
  \end{tabular}
  
  \end{table}

\begin{table}
  \caption{\label{tab:emcand3} Summary of the properties of the 3rd $eeee$ candidate event.  Electron $e_{2}$ does not have a matched track, therefore no charge measurement.}
  \begin{tabular}{lcccc} 
    \hline \hline
    Run Number & \multicolumn{4}{c}{234030} \\ 
    Event Number & \multicolumn{4}{c}{25104195} \\
    \hline
    & $e_1$ & $e_2$ & $e_3$ & $e_4$ \\
    \hline
    $p_T$ (GeV) & $69\pm3$ & $64\pm2$ & $30\pm2$ & $27\pm2$ \\
    $\eta$ & -0.96 & -2.00 & -1.19 & 0.06 \\
    $\phi$ (radians) & 5.41 & 2.54 & 0.05 & 2.53 \\
    $z_{\text{vtx}}$ (cm) & 31.8 & 31.8 & 31.8 & 31.8 \\
    Charge & $-1$ & -- & +1 & +1 \\
    \hline
    &       &        &        & \\ [-3mm]
    & \multicolumn{2}{c}{$e_{1}^{-}e_{4}^{+}$} & 
    \multicolumn{2}{c}{$e_{2}e_{3}^{+}$} \\ [1mm]
    \hline
    Dilepton mass (GeV) & \multicolumn{2}{c}{$98 \pm 3$} & \multicolumn{2}{c}{$90 \pm 3$} \\
    
    $Z/\gamma^* p_{T}$ (GeV) & \multicolumn{2}{c}{$44 \pm 3$} & 
    \multicolumn{2}{c}{$45 \pm 2$} \\
    \hline
    4-lepton mass (GeV) & \multicolumn{2}{r}{239} & \multicolumn{2}{l}{$\pm$ 5} \\
    $Z/\gamma^*Z/\gamma^* p_{T}$ (GeV) & \multicolumn{2}{r}{1}  & \multicolumn{2}{l}{$\pm$ 4} \\
    \Etmiss \ (GeV) & \multicolumn{2}{r}{2} &  \multicolumn{2}{l}{$\pm$ 1} \\
    \hline \hline
  \end{tabular}
  
\end{table}

\begin{table}
  \caption{\label{tab:mucand1} Summary of the properties of the 1st $\mu\mu\mu\mu$ candidate event.}
  \begin{tabular}{lcccc} 
    \hline \hline
    Run Number & \multicolumn{4}{c}{246915} \\ 
    Event Number & \multicolumn{4}{c}{20003687} \\
    \hline
        & $\mu_1$ & $\mu_2$ & $\mu_3$ & $\mu_4$ \\
    \hline
        &       &        &        & \\ [-3mm]
    $p_T$ (GeV) & $56^{+10}_{-7}$ & $49^{+7}_{-6}$ & $41^{+5}_{-4}$ & $38^{+4}_{-4}$ \\ [0.5mm]
    $\eta$ & 0.09 & 0.06 & -0.99 & 0.81 \\
    $\phi$ (radians) & 3.64 & 0.19 & 2.05 & 5.96 \\
    $z_{\text{vtx}}$ (cm) & 15.2 & 15.2 & 15.2 & 15.2 \\
    Charge & +1 & $-1$ & +1 & $-1$ \\
    \hline
    &       &        &        & \\ [-3mm]
    & \multicolumn{2}{c}{$\mu_{1}^{+}\mu_{4}^{-}$} & \multicolumn{2}{c}{$\mu_{2}^{-}\mu_{3}^{+}$} \\ [1mm]
    \hline    
        & \multicolumn{2}{c}{} & \multicolumn{2}{c}{} \\ [-2mm]
    Dilepton mass (GeV) & \multicolumn{2}{c}{$91^{+10}_{-7}$} & \multicolumn{2}{c}{$87^{+8}_{-7}$} \\
      
        & \multicolumn{2}{c}{} & \multicolumn{2}{c}{} \\ [-2mm]
    $Z/\gamma^* p_{T}$ (GeV) & \multicolumn{2}{c}{$41^{+7}_{-5}$} & 
    \multicolumn{2}{c}{$54^{+6}_{-4}$} \\[1mm]
    \hline
    & \multicolumn{4}{c}{} \\[-2mm]
    4-lepton mass (GeV) & \multicolumn{2}{r}{218.5} & \multicolumn{2}{l}{$^{+16}_{-12}$} \\  & \multicolumn{4}{c}{} \\[-2.5mm]
    $Z/\gamma^*Z/\gamma^* p_{T}$ (GeV) & \multicolumn{2}{r}{17} & \multicolumn{2}{l}{$^{+13}_{-10}$} \\ [1mm]
    \Etmiss \ (GeV) & \multicolumn{2}{r}{8} & \multicolumn{2}{l}{$\pm$ 6} \\
    \hline \hline
  \end{tabular}
\end{table}

\begin{table}[t!]

  \caption{\label{tab:mucand2} Summary of the properties of the 2nd $\mu\mu\mu\mu$ candidate event.}
  
  \begin{tabular}{lcccc} 
    \hline \hline
    Run Number & \multicolumn{4}{c}{248990} \\ 
    Event Number & \multicolumn{4}{c}{47671351} \\
    \hline
        & $\mu_1$ & $\mu_2$ & $\mu_3$ & $\mu_4$ \\
    \hline
        &       &        &        & \\ [-3mm]
    $p_T$ (GeV) & $53^{+9}_{-7}$ & $46^{+7}_{-5}$ & $45^{+6}_{-5}$ & $45^{+6}_{-5}$ \\ [0.5mm]
    $\eta$ & -1.12 & -1.96 & -1.14 & -1.78 \\
    $\phi$ (radians) & 4.64 & 2.15 & 2.41 & 5.98 \\
    $z_{\text{vtx}}$ (cm) & 22.4 & 22.4 & 22.4 & 22.4 \\
    Charge & $-1$ & $-1$ & +1 & +1 \\
    \hline
    &       &        &        & \\ [-3mm]
    & \multicolumn{2}{c}{$\mu_{1}^{-}\mu_{3}^{+}$} & 
    \multicolumn{2}{c}{$\mu_{2}^{-}\mu_{4}^{+}$} \\[1mm]
    \hline    
        & \multicolumn{2}{c}{} & \multicolumn{2}{c}{} \\ [-2mm]
    Dilepton mass (GeV) & \multicolumn{2}{c}{$88^{+9}_{-7}$} & \multicolumn{2}{c}{$86^{+8}_{-7}$} \\
       
        & \multicolumn{2}{c}{} & \multicolumn{2}{c}{} \\ [-2mm]
    $Z/\gamma^* p_{T}$ (GeV) & \multicolumn{2}{c}{$44^{+5}_{-4}$} & \multicolumn{2}{c}{$31^{+3}_{-2}$} \\[1mm]
    \hline
      & \multicolumn{4}{c}{} \\[-2mm]
          4-lepton mass (GeV) & \multicolumn{2}{r}{202} & \multicolumn{2}{l}{$^{+15}_{-11}$} \\
    
      & \multicolumn{4}{c}{} \\[-2.5mm]
          $Z/\gamma^*Z/\gamma^* p_{T}$ (GeV) & \multicolumn{2}{r}{20} & \multicolumn{2}{l}{$^{+9}_{-7}$} \\ [1mm]
    
        \Etmiss \ (GeV) & \multicolumn{2}{r}{7} & \multicolumn{2}{l}{$\pm 5$} \\
    \hline \hline
  \end{tabular}
  \end{table}

\begin{table}

  \caption{\label{tab:mucand3} Summary of the properties of the 3rd $\mu\mu\mu\mu$ candidate event.}
  
  \begin{tabular}{lcccc} 
    \hline \hline
    Run Number & \multicolumn{4}{c}{232216} \\ 
    Event Number & \multicolumn{4}{c}{15136574} \\
    \hline
        & $\mu_1$ & $\mu_2$ & $\mu_3$ & $\mu_4$ \\
    \hline
        &       &        &        & \\ [-3mm]
     $p_T$ (GeV) & $116^{+48}_{-26}$ & $78^{+19}_{-13}$ & $42^{+5}_{-4}$ & $24^{+2}_{-2}$ \\ [0.5mm]
    $\eta$ & -0.04 & -1.01 & 0.77 & -1.93 \\
    $\phi$ (radians) & 1.69 & 4.26 & 5.29 & 0.35 \\
    $z_{\text{vtx}}$ (cm) & 17.4 & 17.4 & 17.4 & 17.4 \\
    Charge & +1 & $-1$ & $-1$ & +1 \\
    \hline
    &       &        &        & \\ [-3mm]
    & \multicolumn{2}{c}{$\mu_{1}^{+}\mu_{3}^{-}$} & 
    \multicolumn{2}{c}{$\mu_{2}^{-}\mu_{4}^{+}$} \\[1mm]
    \hline    
        & \multicolumn{2}{c}{} & \multicolumn{2}{c}{} \\ [-2mm]
        Dilepton mass (GeV) & \multicolumn{1}{r}{148} & \multicolumn{1}{l}{$^{+32}_{-18}$} & \multicolumn{2}{c}{$90^{+12}_{-8}$} \\
        
        & \multicolumn{2}{c}{} & \multicolumn{2}{c}{} \\ [-2mm]
            $Z/\gamma^* p_{T}$ (GeV) & \multicolumn{1}{r}{80} & \multicolumn{1}{l}{$^{+47}_{-26}$} & \multicolumn{2}{c}{$62^{+19}_{-12}$} \\[1mm]
    \hline
        & \multicolumn{4}{c}{} \\[-2mm]
                4-lepton mass (GeV) & \multicolumn{2}{r}{347}  & \multicolumn{2}{l}{$^{+55}_{-32}$} \\
    
        & \multicolumn{4}{c}{} \\[-2.5mm]
            $Z/\gamma^*Z/\gamma^* p_{T}$ (GeV) & \multicolumn{2}{r}{18} & \multicolumn{2}{l}{$^{+50}_{-29}$} \\ [1mm]
    
        \Etmiss \ (GeV) & \multicolumn{2}{r}{3}  & \multicolumn{2}{l}{$\pm 2$} \\
    \hline \hline
  \end{tabular}
  \end{table}

\begin{table}

  \caption{\label{tab:mucand4} Summary of the properties of the 4th $\mu\mu\mu\mu$ candidate event.}
  
  \begin{tabular}{lcccc} 
    \hline \hline
    Run Number & \multicolumn{4}{c}{222870} \\ 
    Event Number & \multicolumn{4}{c}{38704512} \\
    \hline
        & $\mu_1$ & $\mu_2$ & $\mu_3$ & $\mu_4$ \\
    \hline
        &       &        &        & \\ [-3mm]
    $p_T$ (GeV) & $63^{+13}_{-9}$ & $34^{+3}_{-3}$ & $22^{+3}_{-2}$ & $20^{+1}_{-1}$ \\ [0.5mm]
    $\eta$ & -1.02 & -1.23 & 1.94 & -1.56 \\
    $\phi$ (radians) & 6.20 & 3.57 & 3.35 & 1.26 \\
    $z_{\text{vtx}}$ (cm) & -16.1 & -16.1 & -16.1 & -16.1 \\
    Charge & +1 & $-1$ & $-1$ & +1 \\
    \hline
    &       &        &        & \\ [-3mm]
    & \multicolumn{2}{c}{$\mu_{1}^{+}\mu_{2}^{-}$} & \multicolumn{2}{c}{$\mu_{3}^{-}\mu_{4}^{+}$} \\ [1mm]
    \hline    
    & \multicolumn{2}{c}{} & \multicolumn{2}{c}{} \\ [-2mm]
    Dilepton mass (GeV) & \multicolumn{2}{c}{$90^{+10}_{-8}$} & \multicolumn{1}{r}{120} & \multicolumn{1}{l}{$^{+9}_{-7}$} \\  
    & \multicolumn{2}{c}{} & \multicolumn{2}{c}{} \\ [-2mm]
    $Z/\gamma^* p_{T}$ (GeV) & \multicolumn{2}{c}{$38^{+12}_{-8}$} & \multicolumn{1}{r}{21} & \multicolumn{1}{l}{$^{+2}_{-1}$} \\ [1mm]
    \hline
    & \multicolumn{4}{c}{} \\[-2mm]
    4-lepton mass (GeV) & \multicolumn{2}{r}{270} & \multicolumn{2}{l}{$^{+22}_{-17}$} \\
    & \multicolumn{4}{c}{} \\[-2.5mm]
    $Z/\gamma^*Z/\gamma^* p_{T}$ (GeV) & \multicolumn{2}{r}{18} & \multicolumn{2}{l}{$^{+13}_{-9}$} \\ [1mm]
    \Etmiss \ (GeV) & \multicolumn{2}{r}{16}  & \multicolumn{2}{l}{$\pm 11$} \\
    \hline \hline
  \end{tabular}
\end{table}

\begin{table}
  \caption{\label{tab:emucand1} Summary of the properties of the 1st $2\mu2e$ candidate event.}
  \begin{tabular}{lcccc} 
    \hline \hline
    Run Number & \multicolumn{4}{c}{244006} \\ 
    Event Number & \multicolumn{4}{c}{24854310} \\
    \hline
        & $e_1$ & $e_2$ & $\mu_1$ & $\mu_2$ \\
    \hline
        &       &        &        & \\ [-3mm]
    $p_T$ (GeV) & $50\pm2$ & $41\pm2$ & $51^{+8}_{-6}$ & $45^{+6}_{-5}$ \\ [0.5mm]
    $\eta$ & 0.52 & -0.76 & -0.03 & -1.30 \\
    $\phi$ (radians) & 2.22 & 0.46 & 3.5 & 5.41 \\
    $z_{\text{vtx}}$ (cm) & 6.5 & 6.5 & 6.5 & 6.5 \\
    Charge & +1 & $-1$ & $-1$ & +1 \\
    \hline
    &       &        &        & \\ [-3mm]
    & \multicolumn{2}{c}{$e_{1}^{+}e_{2}^{-}$} & 
    \multicolumn{2}{c}{$\mu_{1}^{-}\mu_{2}^{+}$} \\ [1mm]
    \hline    
    & \multicolumn{2}{c}{} & \multicolumn{2}{c}{} \\ [-3mm]
    Dilepton mass (GeV) & \multicolumn{2}{c}{$94\pm3$} & \multicolumn{1}{r}{102} & \multicolumn{1}{l}{$^{+11}_{-8}$} \\  
    & \multicolumn{2}{c}{} & \multicolumn{2}{c}{} \\ [-3mm]
    $Z/\gamma^* p_{T}$ (GeV) & \multicolumn{2}{c}{$58\pm2$} & \multicolumn{1}{r}{56.1} & \multicolumn{1}{l}{$^{+6}_{-5}$} \\ [0.5mm]
    \hline
    & \multicolumn{4}{c}{} \\[-2.5mm]
    4-lepton mass (GeV) & \multicolumn{2}{r}{236} & \multicolumn{2}{l}{$^{+13}_{-10}$} \\
    & \multicolumn{4}{c}{} \\[-3mm]
    $Z/\gamma^*Z/\gamma^* p_{T}$ (GeV) & \multicolumn{2}{r}{13} & \multicolumn{2}{l}{$^{+8}_{-7}$} \\ [0.5mm]
    \Etmiss \ (GeV) & \multicolumn{2}{r}{9} & \multicolumn{2}{l}{$\pm 6$} \\
    \hline \hline
  \end{tabular}
  \end{table}

\begin{table}
  \caption{\label{tab:emucand2}Summary of the properties of the 2nd $2\mu2e$ candidate event.}
  \begin{tabular}{lcccc} 
    \hline \hline
    Run Number & \multicolumn{4}{c}{244503} \\ 
    Event Number & \multicolumn{4}{c}{19036212} \\
    \hline
        & $e_1$ & $e_2$ & $\mu_1$ & $\mu_2$ \\
    \hline
         &       &        &        & \\ [-3mm]
    $p_T$ (GeV) & $69\pm3$ & $36\pm2$ & $59^{+11}_{-8}$ & $57^{+10}_{-7}$ \\ [0.5mm]
    $\eta$ & -0.86 & -0.42 & 0.37 & 0.15 \\
    $\phi$ (radians) & 1.59 & 5.27 & 4.76 & 0.14 \\
    $z_{\text{vtx}}$ (cm) & -11.3 & -11.3 & -11.3 & -11.3 \\
    Charge & +1 & $-1$ & $-1$ & +1 \\
    \hline
    &       &        &        & \\ [-3mm]
    & \multicolumn{2}{c}{$e_{1}^{+}e_{2}^{-}$} & 
    \multicolumn{2}{c}{$\mu_{1}^{-}\mu_{2}^{+}$} \\ [1mm]
    \hline    
        & \multicolumn{2}{c}{} & \multicolumn{2}{c}{} \\ [-3mm]
            Dilepton mass (GeV) & \multicolumn{2}{c}{$98\pm3$} & \multicolumn{1}{r}{86} & \multicolumn{1}{l}{$^{+11}_{-8}$} \\
       
        & \multicolumn{2}{c}{} & \multicolumn{2}{c}{} \\ [-3mm]
            $Z/\gamma^* p_{T}$ (GeV) & \multicolumn{2}{c}{$42\pm3$} & \multicolumn{1}{r}{78} & \multicolumn{1}{l}{$^{+10}_{-7}$} \\ [0.5mm]
    \hline
     & \multicolumn{4}{c}{} \\[-2.5mm]
      4-lepton mass (GeV) & \multicolumn{2}{r}{238} & \multicolumn{2}{l}{$^{+15}_{-12}$} \\
        & \multicolumn{4}{c}{} \\[-3mm]
       $Z/\gamma^*Z/\gamma^* p_{T}$ (GeV) & \multicolumn{2}{r}{76} & \multicolumn{2}{l}{$^{+10}_{-7}$} \\ [0.5mm]
    
        \Etmiss  \ (GeV) & \multicolumn{2}{r}{72}  & \multicolumn{2}{l}{$\pm 51$} \\
    \hline \hline
  \end{tabular}
  \end{table}

\begin{table}
  \caption{\label{tab:emucand3}Summary of the properties of the 3rd $2\mu2e$ candidate event..  Electron $e_{1}$ does not have a matched track, therefore no charge measurement.}
  \begin{tabular}{lcccc} 
    \hline \hline
    Run Number & \multicolumn{4}{c}{208914} \\ 
    Event Number & \multicolumn{4}{c}{57115998} \\
    \hline
        &       &        &        & \\ [-3mm]
    $p_T$ (GeV) & $56\pm2$ & $16\pm1$ & $72^{+16}_{-11}$ & $34^{+4}_{-3}$ \\ [0.5mm]
    $\eta$ & -2.13 & 0.61 & 0.80 & 0.25 \\
    $\phi$ (radians) & 5.40 & 6.00 & 3.00 & 1.23 \\
    $z_{\text{vtx}}$ (cm) & 23.3 & 23.3 & 23.3 & 23.3 \\
    Charge & -- & +1 & $-1$ & +1 \\
    \hline
    &       &        &        & \\ [-3mm]
    & \multicolumn{2}{c}{$e_{1}e_{2}^{+}$} & \multicolumn{2}{c}{$\mu_{1}^{-}\mu_{2}^{+}$} \\ [1mm]
    \hline
        & \multicolumn{2}{c}{} & \multicolumn{2}{c}{} \\ [-3mm]
                    Dilepton mass (GeV) & \multicolumn{1}{r}{112} & \multicolumn{1}{l}{$\pm4$} & \multicolumn{1}{r}{79} & \multicolumn{1}{l}{$^{+10}_{-7}$} \\
    
        & \multicolumn{2}{c}{} & \multicolumn{2}{c}{} \\ [-3mm]
                    $Z/\gamma^* p_{T}$ (GeV) & \multicolumn{1}{r}{70} & \multicolumn{1}{l}{$\pm2$} & \multicolumn{1}{r}{76} & \multicolumn{1}{l}{$^{+15}_{-10}$} \\[0.5mm]
    \hline
        & \multicolumn{4}{c}{} \\ [-2.5mm]
            4-lepton mass (GeV) & \multicolumn{2}{r}{359} & \multicolumn{2}{l}{$^{+30}_{-21}$} \\
    
        & \multicolumn{4}{c}{} \\ [-3mm]
            $Z/\gamma^*Z/\gamma^* p_{T}$ (GeV) & \multicolumn{2}{r}{9} & \multicolumn{2}{l}{$^{+15}_{-11}$} \\ [0.5mm]
    
        \Etmiss \ (GeV) & \multicolumn{2}{r}{8} & \multicolumn{2}{l}{$\pm 6$} \\
    \hline \hline
\end{tabular}
\end{table}


\end{widetext}


\begin{thebibliography}{99}
\bibitem{particles}
  Statements concerning particles should also be interpreted to include antiparticles.
\bibitem{pseudo}
     The D0  coordinate system is cylindrical with the 
     $z$-axis along the proton beamline and the polar and azimuthal
     angles denoted as $\theta$ and $\phi$ respectively. The 
     pseudorapidity is defined as $\eta=-\ln[\tan(\theta /2)]$, measured with respect to the event's vertex, where $\eta_{\text{det}}$ is the pseudorapidity measured with respect to the detector's center.
\bibitem{met}  Missing transverse energy \Etmiss~is defined as the opposite of the vector sum of the transverse energies found in the calorimeter.  This \Etmiss~takes into account energy which is carried away by identified muons.
\end{thebibliography}
\end{document}